# Generation and dynamical manipulation of polarization entangled Bell states by a silicon quantum photonic circuit

DONGNING LIU,[1,†] JINGYUAN ZHENG,[1,†] LINGJIE YU,[1,†] XUE FENG,[1] FANG LIU,[1] KAIYU CUI,[1] YIDONG HUANG,[1,2] AND WEI ZHANG[1,2,*]

[1] *Frontier Science Center for Quantum Information, Beijing Innovation Center for Future Chips, Beijing National Research Center for Information Science and Technology (BNRist), Electronic Engineering Department, Tsinghua University, Beijing 100084, China.*
[2] *Beijing Academy of Quantum Information Sciences, Beijing 100193, China.*
*† These authors have equal contributions on this work*
*\* zwei@tsinghua.edu.cn.*

**Abstract:** A silicon quantum photonic circuit was proposed and demonstrated as an integrated quantum light source for telecom band polarization entangled Bell state generation and dynamical manipulation. Biphoton states were firstly generated in four silicon waveguides by spontaneous four wave mixing. They were transformed to polarization entangled Bell states through on-chip quantum interference and quantum superposition, and then coupled to optical fibers. The property of polarization entanglement in generated photon pairs was demonstrated by two-photon interferences under two non-orthogonal polarization bases. The output state could be dynamically switched between two polarization entangled Bell states, which was demonstrated by the experiment of simplified Bell state measurement. The experiment results indicate that its manipulation speed supported a modulation rate of several tens kHz, showing its potential on applications of quantum communication and quantum information processing requiring dynamical quantum entangled Bell state control.

## 1. Introduction

Silicon photonics based on silicon-on-insulator (SOI) platform is a promising route for quantum photonic circuit, which can realize complicated on-chip photonic quantum information systems with large scalability and high stability [1–5]. On one hand, silicon waveguides are good $\chi^{(3)}$ nonlinear media at telecom band. By spontaneous four wave mixing (SFWM), high-quality correlated biphoton state could be generated in silicon waveguides of several millimeters [6-8]. The generation efficiency can be further improved if resonances in devices composed of silicon waveguides are used, such as micro-ring resonators [9–12]. On the other hand, silicon photonics based on SOI platform is compatible with the complementary metal oxide semiconductor (CMOS) technology [13,14]. Large-scale photonic circuits for complicated quantum state manipulation could therefore be supported [15–17]. Hence, it is a promising way to develop telecom-band quantum light sources for various biphoton state generation [18,19].

Polarization entangled biphoton states have been widely used in experiments of quantum optics and quantum information [20], since it is easy to be manipulated by linear optics. Accordingly, it is crucial to develop quantum light sources for polarization entangled biphoton state generation. Traditionally, these biphoton states are generated by spontaneous parametric down conversion (SPDC) in nonlinear crystals [21,22]. To realize telecom band integrated quantum light sources, several schemes based on SFWM in silicon waveguides have been proposed and demonstrated [7,23,24]. In silicon waveguides, the fundamental quasi-transverse electrical (quasi-TE) mode is preferred in the biphoton state generation, since the efficiency of SFWM in this mode is much higher than those in other modes [25]. Hence, specific designs are required to realize the quantum superposition of biphoton states with orthogonal polarization directions, such as a modified Sagnac loop [7], a polarization rotator embedded in the middle of the silicon waveguide [23], or two-dimensional gratings [24]. However, in these schemes, only one mono-color pump light is used, and the signal and idler photons have different frequencies. Hence, the generated states are not polarization entangled Bell states, which should be discriminated by Bell state measurement (BSM) [26]. Usually, BSM is realized by setups based on linear optics and the quantum interference in them requires that the two input photons should be indistinguishable except the freedom of polarization. Hence, in a photon pair with a polarization entangled Bell state, the two photons should be indistinguishable in their frequencies. On the other hand, silicon photonics provides convenient ways to realize on-chip quantum state manipulation [4, 27]. However, the performance of dynamical manipulation on quantum states in silicon quantum photonic circuits has not been demonstrated, including the experiments of these quantum light sources based on silicon waveguides.

In this work, we proposed and demonstrated a silicon quantum photonic circuit to realize integrated quantum light source for telecom-band polarization entangled Bell state generation. It was based on SFWM in the fundamental quasi-TE modes of four silicon waveguides. Stimulated by two pump lights with different frequencies, two pump photons with different frequencies (denoted by $\omega_{p1}$ and $\omega_{p2}$) were annihilated, while two photons with the same frequency (denoted by $\omega_{s,i}$) were generated simultaneously through this SFWM process. The energy conservation was satisfied in SFWM, leading to the relation of $\omega_{p1}+\omega_{p2}=2\omega_{s,i}$. The biphoton states generated in the four silicon waveguides were transformed to polarization entangled Bell states by on-chip quantum interference and quantum superposition, and then coupled to optical fibers. By adjusting the phases of the quantum interference and quantum superposition, the output state could be dynamically switched between two polarization entangled Bell states by electrical signal. The performance of the dynamical manipulation was measured to show its potential on applications requiring dynamical Bell state control.

## 2. The silicon quantum photonic circuit

The silicon quantum photonic circuit was fabricated on a SOI substrate with a silicon layer of 220 nm in thickness. Its sketch is shown in Fig. 1. It has six input/output ports, denoted by Port 1~6, respectively. Two pump lights with different frequencies are injected into the photonic circuit through Port 1 and 2, respectively. Both of them are coupled to the fundamental quasi-TE mode of the silicon waveguides by the gratings on silicon waveguides (denoted by WG). Then they are split to four long silicon waveguides (denoted by W1~4, respectively) through three 50:50 beam splitters (denoted by BS 1~3, respectively). All the 50:50 beam splitters on the photonic circuit are realized by 2×2 multi-mode interference (MMI) devices. The four long silicon waveguides are used as the nonlinear media of SFWM with the same length of 1 cm. When pump lights propagate along these waveguides, biphoton states are generated by SFWM. The biphoton states generated in W1 and W2 interfere at the fourth 50:50 beam splitter (BS4). The phase of this interference can be adjusted by a thermal phase shifter (TPS2) on W1. The output state of this quantum interference is a superposition of two biphoton states denoted as $|\Psi_{bunch}\rangle$ and $|\Psi_{split}\rangle$, respectively [28]. For the photon pairs of $|\Psi_{bunch}\rangle$, the two photons always output from the same port, but which port they select is random. On the contrary, for the photon pairs of $|\Psi_{split}\rangle$, the two photons always output from two different ports, respectively. To realize the polarization entangled Bell state generation, the output state should be adjusted to $|\Psi_{split}\rangle$ by controlling the interference phase through TPS2. This state can be expressed as $|\varphi\rangle_{BS4} = |c, d\rangle_{BS4}$, where "c" and "d" denote the two output ports of BS4, respectively. Similarly, the biphoton states generated in W3 and W4 interfere at the fifth beam splitter (BS5), with proper interference phase control by another thermal phase shifter (TPS3) on W3. The output state of BS5 also can be adjusted to $|\Psi_{split}\rangle$, which can be expressed as $|\varphi\rangle_{BS5} = |c', d'\rangle_{BS5}$, where "c'" and "d'" denote the two output ports of BS5, respectively.

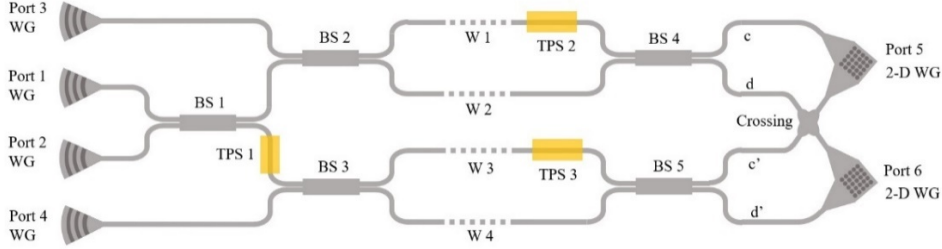

**Fig. 1** The sketch of the silicon quantum photonic circuit for polarization entangled Bell state generation and dynamical manipulation. WG: grating on silicon waveguide; BS: beam splitter; TPS: thermal phase shifter; W: long silicon waveguide; 2-D WG: two-dimensional grating on silicon waveguide

The output ports of BS4 and BS5 are connected to two two-dimensional gratings on silicon waveguides (2-D WGs) as shown in Fig. 1. These two 2-D WGs are two output ports of the photonic circuit (Port 5 and 6), by which the generated photon pairs are coupled to two optical fibers. The photons in the two input waveguides of the 2-D WGs would be coupled into two orthogonal polarization modes of the output fiber [29,30], which are denoted by "H" and "V", respectively. By this way, the freedom of paths on the photonic circuit are transformed into the freedom of polarization directions in the two output fibers. The relation between them is defined as Table 1.

**Table 1. The relation between the output ports of BS 4/5 and the polarization directions in output fibers**

| Output ports of beam splitters | Output ports of the photonic circuit | Polarization directions in fibers |
|---|---|---|
| "c" of BS4 | Port 5 | H |
| "d" of BS4 | Port 6 | V |
| "c'" of BS5 | Port 5 | V |
| "d'" of BS5 | Port 6 | H |

According to this relation, the output biphoton state in the two optical fibers is

$$|\Psi\rangle = \frac{1}{\sqrt{2}}(|H,V\rangle + e^{i\alpha}|V,H\rangle) \quad (1)$$

It is a maximum polarization entangled state. α is the phase of the quantum superposition, which can be adjusted by the thermal phase shifter (TPS1) between BS 1 and BS 3. Particularly, if α is adjusted to 0 or π, two polarization entangled Bell states can be generated as follows,

$$|\Psi^{\pm}\rangle = \frac{1}{\sqrt{2}}(|H,V\rangle \pm |V,H\rangle) \quad (2)$$

These two biphoton states can be discriminated by BSM based on linear optics.

It is worth noting that Port 3 and 4 are auxiliary input ports connected to BS 2 and BS 3, respectively. They are used to calibrate the thermal phase shifters on the photonic circuit and the experimental setup for the measurement of the output biphoton states. In the measurement, the frequency-degenerate photon pairs would be selected by optical filters before the single photon detectors.

## 3. The experiment results

### 3.1 The indistinguishability of the two photons in photon pairs

In the experiment, the wavelengths of the two pump lights were 1555.7 nm and 1549.3 nm, respectively. Firstly, they were combined and injected into Port 3, stimulating SFWM in waveguides W1 and W2. The generated biphoton states interfered at the 50:50 beam splitter

BS4. The phase of this interference was controlled by the thermal phase shifter TPS2. The output photons from Port 5 and 6 were coupled to two optical fibers. The generated frequency-degenerate photon pairs were selected by optical filters and detected by two single photon detectors. The central wavelength and full width at half maximum (FWHM) of the optical filters were 1552.5 nm and ~60 GHz, respectively. Their sideband suppression ratios were both over 100 dB to eliminate the impact of pump lights. The single photon detectors were based on near infrared avalanche photo diodes with efficiencies of ~20% and dark count rates of ~100 Hz. Figure 2 (a) shows the coincidence counts of the two single photon detectors under different voltages on TPS2 (It is indicated by the square of the voltage due to Ohm's law). It can be seen that the coincidence counts vary sinusoidally with the square of the voltage, showing the fringe of the quantum interference. It agrees with the theoretical analysis that the output state after BS4 is a superposition of $|\Psi_{bunch}\rangle$ and $|\Psi_{split}\rangle$, and the phase of the superposition could be controlled by TPS2. Especially, indicated by the red arrow in Fig. 2(a), the coincidence counts reached its maximum when the voltage was 7.47V. It is the condition that the output state only includes $|\Psi_{split}\rangle$ and the two photons in a pair output from Port 5 and 6 respectively.

Under this condition, we took the experiment of HOM interference to test the indistinguishability between the photons in generated photon pairs. A 50:50 fiber coupler was connected to the two output fibers. The coincidence counts of the output photons from the fiber coupler were detected by single photon detectors with the optical filters. At one input port of the fiber coupler, a fiber polarization controller was placed to maximize the quantum interference by making the input photons indistinguishable in polarization. At the other input port of the fiber coupler, a variable delay line was placed to adjust the time delay between the two photons in a pair. Figure 2(b) shows the coincidence counts under different time delays. It can be seen that the fringe shows a clear dip when the time delay is close to zero, which is the feature of HOM interference. The red line in the figure is the fitting curve of the HOM dip, under the assumption that the optical filters for the frequency-degenerate photon pairs have rectangular transmission spectra. The raw visibility of the HOM dip is 91.0%. Similarly, we also measured the HOM interference of the photons generated in waveguide W3 and W4 by injecting the pump lights into Port 4. The results are shown in Fig. 2(c). The fringe also shows the clear feature of HOM interference with a raw visibility of 93.9%. These results show that the photons of the generated photon pairs have good indistinguishability, especially in their spectral properties.

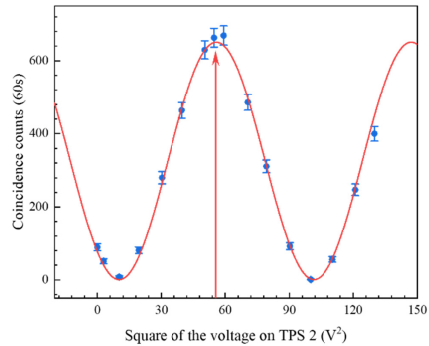

(a)

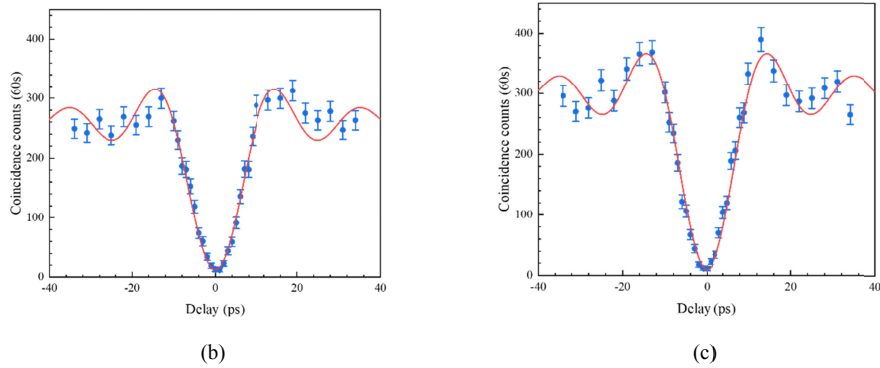

(b)                                      (c)

**Fig. 2** The experiment to show the indistinguishability of the two photons in photon pairs (a) Results of quantum interference between the biphoton states generated in W1 and W2 at BS4. (b) Results of HOM interference between photons in frequency-degenerate photon pairs generated in W1and W2. (c) Results of HOM interference between photons in frequency-degenerate photon pairs generated in W3 and W4

## *3.2 The polarization entanglement of the generated photon pairs*

The property of polarization entanglement was demonstrated by the experiment of two-photon interference under two non-orthogonal polarization bases. In this experiment, the two pump lights were injected into the photonic circuit from Port 1 and Port 2, respectively. The biphoton states after BS 4 and BS 5 were adjusted to $|\Psi_{split}\rangle$ by controlling the voltages on TPS2 and TPS3. Under this condition, the frequency-degenerate polarization entangled biphoton state as Eq. (1) would be generated and coupled to two optical fibers from Port 5 and Port 6. The experiment setup is shown in Fig. 3(a). The output photons in the two optical fibers passed through the polarization analyzer and then detected by the single photon detectors with the optical filters, respectively. The polarization analyzer included fiber polarization controllers (FPC1 and FPC2), rotated half-wave plates (HWP1 and HWP2) and fixed polarizers (P1 and P2). The rotated half-wave plates and the fixed polarizers were mounted on benches with fiber collimators. It was collimated by a probe light at 1552.5 nm before the experiment. We measured the single side counts and coincidence counts of the two single photon detectors under varying direction of HWP2 when the direction of HWP1 was set at 0º and 22.5º, respectively. The results of single side counts are shown in Fig. 3(b). It can be seen that the single side counts have small fluctuations with varying direction of HWP2. It is due to the polarization-dependent loss (PDL) of the 2-D WGs we used and could be removed by more complicated 2-D WG designs developed recently [30]. The results of coincidence counts are shown in Fig. 3 (c). The raw visibilities of the two sinusoidal fringes are 89.5% and 77.7%, respectively, higher than the criterion of violation of Bell inequality [21]. Hence, the photons in generated photon pairs are entangled in polarization.

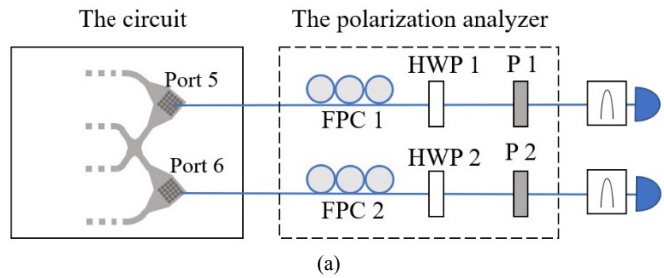

(a)

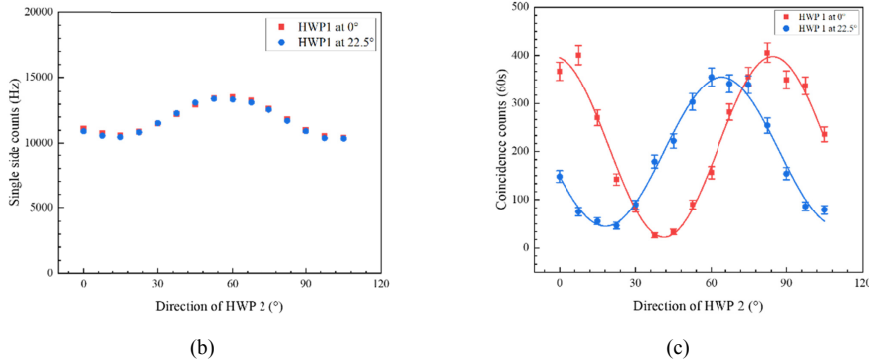

(b)                                                             (c)

**Fig. 3** Experiment results of two photon interference under two non-orthogonal polarization bases. (a) Single side counts and (b) coincidence counts under varying direction of HWP 2 when the direction of HWP 1 was set at 0º and 22.5 0º, respectively.

### *3.3 Two polarization Bell states generated by the photonic circuit*

The experiment results in Fig. 3 show that the photonic circuit generates telecom-band polarization entangled biphoton state as Eq. (1). In the state, the phase α could be controlled by the thermal phase shifter TPS 1. Hence, two polarization entangled Bell states could be generated under α=0 and α=π, which are shown in Eq. (2). They could be demonstrated by a simplified BSM based on a 50:50 fiber coupler. For the photon pairs of $|\Psi^-\rangle$, the two photons in a pair would output from two different ports of the 50:50 fiber coupler respectively, leading to one coincidence count. For the photon pairs of $|\Psi^+\rangle$, the two photons in a pair would output from the same port. Hence, no coincidence count would be recorded. In the experiment, the two output fibers of the photonic circuit connected to the two input ports of the fiber coupler through a variable delay line and a fiber polarization controller, respectively. The output photons from the fiber coupler were detected by the two single photon detectors with the optical filters. Firstly, we optimized the voltage on TPS1 and the settings of the variable delay line and the fiber polarization controller to achieve the minimum coincidence counts. It was the feature of $|\Psi^+\rangle$. Then, we measured the coincidence counts under varying voltage on TPS1. The result is shown in Fig. 4(a). It is a sinusoidal curve varying with the square of the voltage on TPS1. The blue arrow and the red arrow indicate the voltage for the minimum and maximum coincidence counts on the fringe, which are the conditions to generate $|\Psi^+\rangle$ and $|\Psi^-\rangle$, respectively. We set the voltage on TPS 1 at these conditions and measure the coincidence counts under different time delays between the two photons in a pair by tuning the variable delay line in both cases. The results are shown in Fig. 4(b). The red hollow squares and blue solid circles in the figure are the results of $|\Psi^+\rangle$ and $|\Psi^-\rangle$, respectively. It can be seen that they are almost the same when the time delay is large, since no quantum interference took place in this condition. On the other hand, a significant difference between them appears when the time delay is close to zero, showing that the two biphoton states can be discriminated by the simplified BSM. The raw visibility of the two fringes is 87.2%, which is calculated by

$$F = \frac{C_{\max} - C_{\min}}{C_{\max} + C_{\min}} \qquad (3)$$

Where $C_{\max}$ and $C_{\min}$ denote the maximum coincidence count for $|\Psi^-\rangle$ and minimum coincidence count for $|\Psi^+\rangle$, respectively. These results show that two polarization entangled Bell states, $|\Psi^+\rangle$ and $|\Psi^-\rangle$, were generated successfully by the electrical control on the photonic circuit. They can be discriminated by BSM based on linear optics with good fidelity.

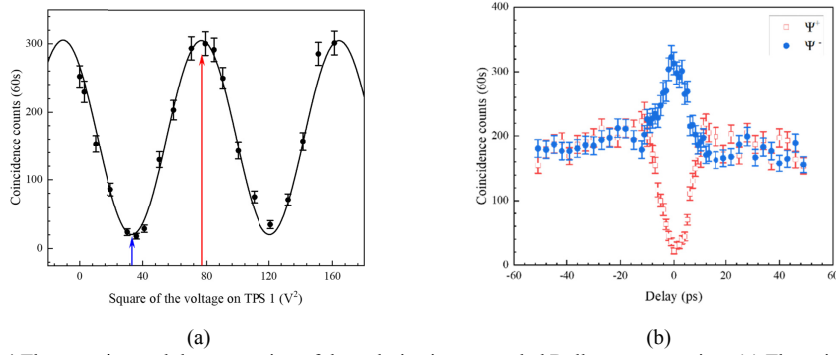

**Fig. 4** The experimental demonstration of the polarization entangled Bell state generation. (a) The coincidence counts under different voltages on TPS1. The blue and red arrows indicate the voltages for $|\Psi^+\rangle$ and $|\Psi^-\rangle$, respectively. (b)The coincidence counts under different time delays of the two photons in a pair at the 50:50 fiber coupler. The results of $|\Psi^+\rangle$ and $|\Psi^-\rangle$ are shown by the red hollow squares and the blue solid circles, respectively.

### *3.4 Dynamical manipulation of the two polarization entangled Bell states*

Dynamical manipulation of polarization entangled Bell states has important applications on quantum communication and quantum information processing requiring quantum encoding, such as quantum dense coding and quantum secure direct communication, etc. The output biphoton state of the silicon quantum photonic circuit in this work can be switched dynamically between $|\Psi^+\rangle$ and $|\Psi^-\rangle$. We demonstrated this function by the simplified BSM. A square wave signal is applied on TPS1, the voltages in the first and second halves of a period were set at the voltage for $|\Psi^+\rangle$ and $|\Psi^-\rangle$, respectively. The coincidence counts of the output photons from the fiber coupler were measured under the optimized condition by controlling the variable delay line and the fiber polarization controller. A period of the square wave was split into 40 time bins, and coincidence events in different periods were counted in these time bins. The histograms shown in Fig. 5 (a) and (b) are the experimental results when the repetition rates of the square wave signal were 1kHz and 20kHz, respectively. In these figures, each bar indicates the coincidence counts in a specific bin. It can be seen that the coincidence counts in the first and second halves of a period have significant difference under a repetition rate of 1kHz, indicating that the output biphoton state was switched between $|\Psi^+\rangle$ and $|\Psi^-\rangle$ by the square wave signal successfully. When the repetition rate increases to 20kHz, the transient process of the biphoton state manipulation can be observed. It shows that the photonic circuit supports a dynamical biphoton state modulation of several tens of kHz. The manipulation speed is typical for the silicon photonic circuits with thermal phase shifters [31,32]. To our knowledge, it is the first demonstration on the on-chip dynamical manipulation of entangled biphoton states. It is worth noting that higher modulation speed could be expected if phase shifters based on electro-optic effect are used in this photonic circuit design. Recent developments of hybrid silicon and lithium niobate devices show great potential on this application [33].

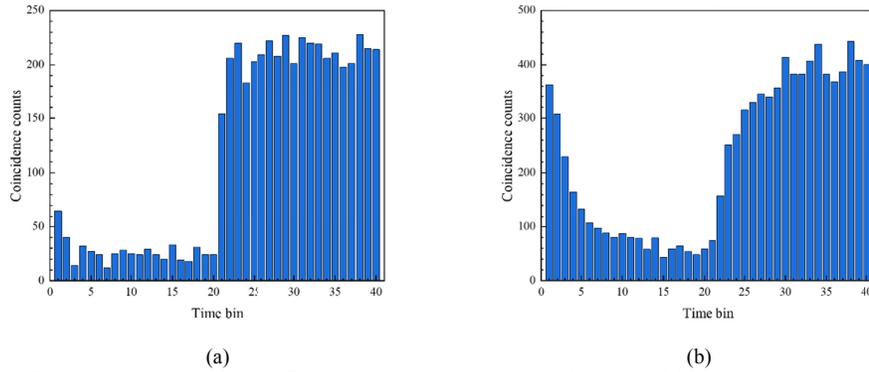

(a)　　　　　　　　　　　　　　(b)

**Fig. 5** The dynamical manipulation of the two polarization entangled Bell states. (a) the results under a square wave repetition rate of 1kHz and a counting time of 600s. (b) the results under a square wave repetition rate of 20kHz and a counting time of 1200s.

## 4. Conclusions

In this work, generation and dynamical manipulation of polarization entangled Bell states were realized by a silicon quantum photonic circuit. In the photonic circuit, biphoton states were generated in four long silicon waveguides by SFWM. They were transformed to polarization entangled Bell states by quantum interference and quantum superposition, and then outputted through optical fibers. The property of polarization entanglement in the output states was demonstrated by two photon interferences under two non-orthogonal polarization bases, with raw fringe visibilities of 89.5% and 77.7%, respectively. The output state can be switched between two polarization entangled Bell states, $|\Psi^+\rangle$ and $|\Psi^-\rangle$, through the electrical signal applied on thermal phase shifters. It was demonstrated by a simplified BSM experiment, showing a visibility of 87.2% in the coincidence measurements of $|\Psi^+\rangle$ and $|\Psi^-\rangle$. By applying square wave signal on the thermal phase shifter, the dynamical process of the Bell state modulation was observed. It showed that the silicon quantum photonic circuit supports a modulation rate of several tens kHz. As a telecom band quantum light source with the function of dynamical quantum state manipulation, the photonic circuit is promising on applications of quantum communication and quantum information processing.

**Funding.** National Key R&D Program of China (2017YFA0303704, 2018YFB2200400), Natural Science Foundation of Beijing (Z180012), National Natural Science Foundation of China (61875101, 91750206), Beijing Academy of Quantum Information Science (Y18G26), Tsinghua Initiative Scientific Research Program.

**Disclosures.** The authors declare no conflicts of interest.

**Data availability.** Data underlying the results presented in this paper are not publicly available at this time but may be obtained from the authors upon reasonable request.